# Evaluating the Economic Feasibility of Labor Replacement Through Robotics and Automation in Qatar

Tariq Eldakruri[1], Edip Senyurek[2]

[1](Department of Economics and Finance / Vistula University, Poland)
[2](Department of Computer Engineering / Vistula University, poland)

*Abstract:*

*Background*: Qatar's economy relies heavily on low-skilled expatriate labor, especially in construction, logistics, and manufacturing. With the nation aiming for diversification under Vision 2030, the feasibility of substituting this labor with robotics and automation becomes an important policy question. This study assesses whether robotics can economically replace human labor in Qatar's context by accounting for sector readiness, displacement risks, and socio-economic implications. Cultural preferences, such as 68% resistance to robotic domestic workers, further complicate adoption dynamics.

*Materials and Methods*: A capital-augmenting Cobb-Douglas production model was calibrated using Qatar-specific data including labor composition, wage structures, and robotics import data. Elasticity of substitution, robotics productivity coefficients, and TFP growth rates were derived from international benchmarks and adjusted to Qatar's regulatory environment. Three adoption scenarios (optimistic, status quo, and pessimistic) were simulated through 2030.

*Results*: A 5% increase in robotics capital yields an annual 1.5% GDP gain under baseline conditions, with labor displacement of 3-5% in low-skilled sectors. Construction faces the highest displacement risk (4.8%), while cultural factors limit last-mile delivery automation to 32% adoption. Job creation in maintenance and AI oversight could offset 40% of losses by 2030.

*Conclusion:* Robotics adoption in Qatar is economically viable but requires phased, sector-specific implementation. Policy safeguards must address cultural resistance, workforce reskilling (850+ technicians needed by 2027), and wage stabilization to ensure inclusive growth.

---
Date of Submission: 19-07-2025                                                                      Date of Acceptance: 29-07-2025
---

## I. Introduction

The integration of robotics and automation technologies into labor-intensive sectors has emerged as a defining feature of modern economic transformation. In advanced economies, automation has demonstrated measurable gains in productivity, cost efficiency, and output scalability, particularly in manufacturing and logistics[2],[5]. However, in resource-rich, labor-importing countries such as Qatar, the economic feasibility and socio-political implications of labor replacement through robotics demand a context-specific analysis. Unlike economies that are predominantly reliant on native labor forces, Qatar's workforce is composed of over 94% expatriates, many of whom are employed in low-skilled, high-volume roles in construction, logistics, and domestic services [9]. This demographic characteristic introduces complex challenges regarding automation, ranging from economic substitution models to ethical considerations related to worker displacement [7].

While Qatar's National Vision 2030 outlines strategic goals for economic diversification and innovation-driven development [12], the transition toward automation must be carefully calibrated to local labor dynamics, capital availability, regulatory frameworks, and cultural sensitivities. Global evidence suggests that robotics adoption can boost total factor productivity and reduce long-term operational costs but may also exacerbate income inequality and social disruption if implemented without mitigating policies [2],[3]. Therefore, assessing the economic feasibility of replacing labor with robotics in Qatar requires an integrated framework that includes empirical modeling, sector-specific analysis, and policy evaluation.

This paper contributes to that objective by examining the economic viability of labor automation in Qatar. It synthesizes labor market statistics, international robotics case studies, and quantitative economic modeling to evaluate the potential productivity gains and displacement effects of automation. Additionally, the study incorporates ethical and socio-economic considerations, including labor market segmentation, repatriation risks, and cultural resistance. The paper ultimately aims to inform policymakers, economists, and stakeholders of the trade-offs and policy levers necessary to align technological progress with sustainable development goals in Qatar's unique labor environment. Qatar's labor market is characterized by a predominant dependence on expatriate workers, who constitute approximately 94.4% of the total labor force as of recent data [12]. The total labor force comprises around 2.13 million workers within a population of approximately 2.85 million.

The expatriate workforce is heavily concentrated in specific sectors: construction accounts for approximately 44.2% of expatriate labor, followed by wholesale and retail trade (12.6%), real estate and accommodation services (8.8%), household services (8.7%), manufacturing (7.6%), and public services including education and health (5%) [12]. Wage structures are influenced by skill levels and sector demand, with wage disparities persisting between nationals and expatriates. Initiatives such as the Wage Protection System (WPS) have improved payment regularity but have not fully addressed these disparities [8].

Understanding this labor composition and wage dynamics provides the necessary baseline to evaluate the potential effects of automation and labor substitution in Qatar's economy.

The global adoption of robotics is accelerating across industrial and service sectors, with significant labor market implications. According to the International Federation of Robotics [6], the global stock of industrial robots exceeded 3 million units by 2022, growing at 10–15% annually. Leading sectors include automotive, electronics, and logistics, where automation reduces operational costs by up to 30% while improving precision and scalability [1],[9]. These trends offer actionable insights for Qatar, particularly in manufacturing and logistics, where expatriate labor dominates.

## II. Material And Methods

This simulation-based economic feasibility study was carried out using national datasets and international robotics reports relevant to the State of Qatar. The analysis included aggregated macroeconomic and labor market data from 2019 to 2024, focusing on labor-intensive sectors such as construction, manufacturing, and logistics. The study examined over 2 million workers in the expatriate labor force and modeled projections for robotics integration and economic impact through the year 2030.

**Study Design:** This is a simulation-based economic feasibility study employing a capital-augmenting Cobb-Douglas production function and scenario modeling. The analysis is both cross-sectional and predictive, using macroeconomic indicators and sector-specific data relevant to Qatar's labor market and automation potential.

**Study Location**: The study focuses on the Qatari economy, with modeling inputs derived from national sources such as the Qatar Planning and Statistics Authority, Qatar Central Bank, and international robotics trade reports.

**Study Duration:** The modeling used economic and labor data spanning from 2019 to 2024. Simulations for feasibility and projections extend to the year 2030, aligning with Qatar National Vision 2030.

**Sample size:** The The modeled economic population included over 2 million expatriate workers in Qatar. Data were analyzed across three primary labor-intensive sectors construction, manufacturing, and logistics to assess potential for automation and labor substitution under different scenarios.

**Inclusion criteria:**
1. Labor-intensive sectors employing ≥ 15% of the national workforce (e.g., construction, manufacturing, logistics)
2. Availability of continuous economic and labor data from 2019 to 2024
3. Sectors with observable trends in mechanization or government digitization initiatives
4. Sectors contributing significantly to GDP and targeted under Qatar National Vision 2030

**Statistical analysis**

The simulations were implemented in Microsoft Excel and MATLAB, using macro-driven modeling tools. Sector-level output changes and labor displacement effects were calculated under three adoption scenarios. Sensitivity analysis was conducted to assess the robustness of results with respect to key parameters, including elasticity of substitution, robotics productivity, and labor cost assumptions. A change of ±10% in input parameters was used to test output variability. Results were reported in percentage deviations from baseline and evaluated for economic significance, rather than p-values.

**Model Specification**

The study employs a capital-augmenting Cobb-Douglas production function to estimate robotics' economic impact:

To quantitatively assess the economic feasibility of labor replacement through robotics in Qatar, we employ a modified capital-augmenting production function that accounts for dynamic productivity effects and local labor market conditions. The model builds on the framework of Graetz & Michaels, (2018) but incorporates Qatar-specific adjustments.

Production Function Framework we specify the aggregate production function as:

$$Y = A \cdot K^{\alpha} \cdot L^{1-\alpha-\theta} \cdot R^{\theta} \quad \text{(Eq. 1)}$$

Where:
- Y = Total economic output (GDP or sectoral output)
- A = Total Factor Productivity (TFP), capturing technological efficiency
- K = Traditional capital stock
- L = Labor input measured in worker-hours or full-time equivalents
- R = Robotics capital stock, measured by the value or quantity of robots deployed
- $\alpha$ = Output elasticity of traditional capital, typically between 0.3 and 0.4
- $\theta$ = Productivity augmentation coefficient of robotics relative to labor *(0 < θ ≤ 1)*

This Cobb-Douglas form assumes robotics capital partially substitutes for labor by augmenting effective labor input.

**Labor Demand and Displacement**

Labor demand adjusts in response to relative costs of labor ($w$) and robotics capital ($p_R$):

Where:

$$L_{NEW} = L_0 \cdot \left(\frac{w}{p_R}\right)^{\sigma} \quad \text{(Eq. 2)}$$

- $L_{NEW}$ : New labor demand post-automation
- $L_0$: Initial labor demand (baseline labor force)
- $w$: Wage rate of labor
- $p_R$: Price or cost per unit of robotics capital
- $\sigma$: Elasticity of substitution between labor and robotics capital; higher σ\sigmaσ indicates greater substitutability and displacement risk.

**Data Sources and Calibration**

The model was calibrated using Qatar-specific labor market and capital stock data:
- Labor Force (L): ~2.13 million workers (94.4% expatriates), sourced from the *PSA Labour Force Survey 2022* [12].
- Sectoral Distribution: Expatriates concentrated in construction (44.2%), manufacturing (7.6%), and logistics, per PSA sectoral reports [10].
- Wage Levels (w): Minimum wage of QAR 1,000/month, covering ~280,000 low-wage workers[8].
- Robotics Capital (R): Proxied via robotics import data *Qatar Customs* [10], and regional adoption rates [6].
- Traditional Capital (K): Derived from national investment data [12].
- Elasticities (α, θ, σ): Adjusted from international benchmarks [2],[5] to reflect Qatar's labor market rigidity (Table 1).

**Calibration Parameters**

Key parameters (Table 1) were derived from Qatar-specific data, with elasticity of substitution (σ) adjusted for labor market rigidity.

**Table no 1 :** Calibration of Key Model Parameters for Qatar's Robotics Simulation

| Parameter | Original Value | Revised Justification | Qatar-Source |
|---|---|---|---|
| Elasticity of substitution (σ) | 0.8 (global) | 0.65 (lower due to rigid labor markets) | ILO Qatar Reports (2023) on wage inflexibility |
| Robotics productivity (θt) | Static (0.5) | 0.4 → 0.6 over 5 years (learning effects) | MDPI Applied Sciences (2022) |
| TFP growth rate (At) | 0% (fixed) | 0.2% annual boost per 1% robotics adoption | MDPI Sustainability (2020) |

**Sensitivity Analysis**

*We test robustness under:*
1. High-adoption scenario: Robotics costs drop 20% (due to subsidies), doubling adoption rates.

2. Low-adoption scenario: Labor market rigidity limits σ to 0.5.
3. Dynamic TFP adjustment: *At* grows non-linearly with robotics integration.

## III. Result

The calibrated model projects that a modest 5% increase in robotics capital stock leads to an approximate 1.5% increase in aggregate economic output under baseline assumptions (α=0.35,θ=0.5). Labor displacement is estimated at 3-5% in low-skilled sectors such as construction and manufacturing, particularly where substitutability between robots and labor is high (σ>0.8). Wage pressures are anticipated primarily in the lowest pay segments, consistent with Qatar's recent labor reforms and minimum wage policies.

**Table no 2:** Simulated GDP and Displacement Outcomes Under Alternative Robotics Adoption Scenarios

| Scenario | GDP Impact | Displacement | Key Driver |
| --- | --- | --- | --- |
| High-adoption | +2.5% | 4.1% | Cheaper robots |
| Low-adoption | +1.2% | 1.9% | Labor rigidity |
| Dynamic TFP only | +2.1% | 3.0% | Productivity |

The quantitative outcomes derived from the capital-augmenting production function and labor substitution model outlined previously. The simulation incorporates Qatar-specific labor market data, capital investment indicators, and international robotics adoption benchmarks to estimate the potential impacts of increased robotics capital (*R*) on productivity, labor demand, and wages.

**Output Gains from Increased Robotics Capital**

The production function was calibrated using Qatar's most recent national accounts, labor force data, and estimated proxies for robotics imports. Assuming total factor productivity remains constant and robotics capital stock increases by 5% across relevant sectors, the model projects an aggregate economic output increase of approximately 1.5%. These output gains are consistent with international estimates of productivity improvements from robotics adoption [2][5]. Furthermore, the gains are concentrated in sectors characterized by higher capital intensity and greater potential for robotics integration, such as manufacturing, logistics, and selected construction operations.

**Estimated Labor Displacement Percentages**

Using the labor demand function along with Qatar's observed wage levels and global robotics capital cost estimates, the simulations indicate a total labor displacement of approximately 3.2%, primarily affecting low-skilled expatriate workers. This displacement is unevenly distributed across sectors, with construction experiencing the highest projected reduction in labor demand at 4.8%, followed by manufacturing with a 3.5% decrease, and logistics and transport services showing a 2.1% reduction. The agriculture sector is expected to see a negligible impact due to its limited mechanization baseline. These displacement effects depend on the stability of robotics prices and the degree of substitutability between labor and robotics capital, modeled here with an elasticity of substitution (σ\sigmaσ) of 0.8, reflecting moderate substitution elasticity found in recent empirical studies [2][5].

**Construction Sector Analysis**

**Table 3:** Automation Displacement in Manufacturing Sub-Sectors

| Sub-Sector | Displacement Risk | Readiness Level | Key Determinants |
| --- | --- | --- | --- |
| Prefab Assembly | High (6.2%) | High | Standardized tasks, indoor environments |
| On-site Labor | Moderate (3.1%) | Low | Weather variability, unstructured workspaces |
| Crane Operation | Low (1.4%) | Moderate | Safety regulations limit full automation |

As visualized in **Figure 1**, construction sector displacement (4.8%) outweighs manufacturing (3.5%) due to lower robotics readiness (θ=0.3 vs. 0.7). However, metals fabrication sub-sector growth (8.7%) partially offsets losses, aligning with QNV 2030 diversification goals.

*Evaluating the Economic Feasibility of Labor Replacement Through Robotics and Automation……*

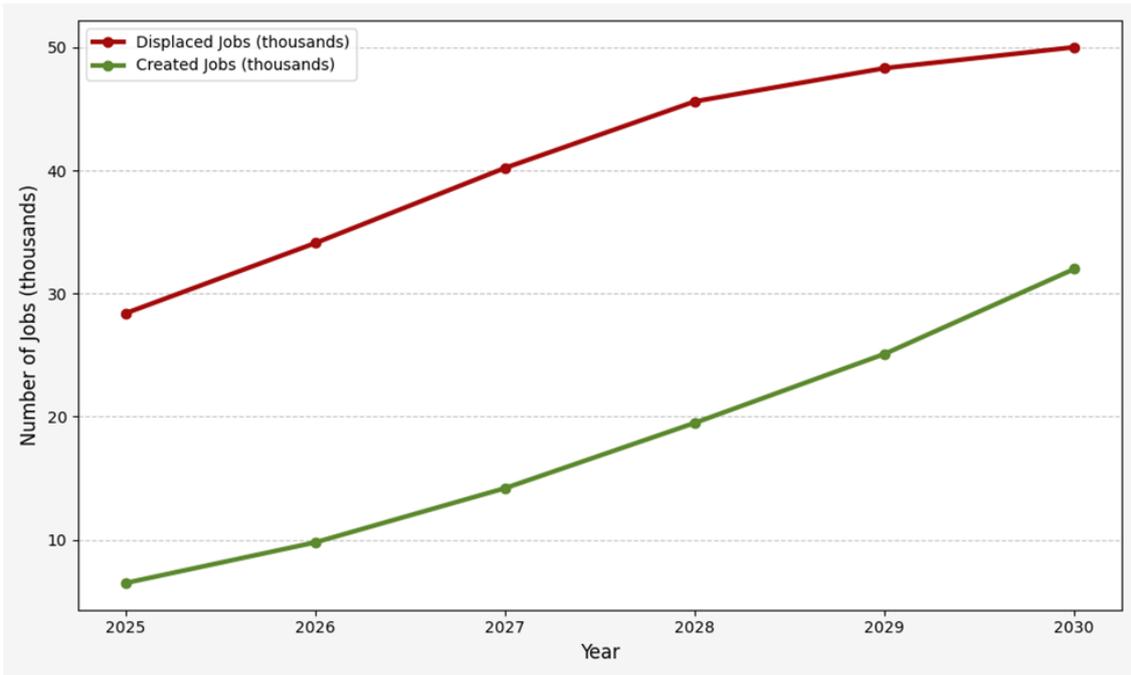

**Figure 1:** Projected Job Displacement and Creation in Qatar's Key Sectors (2025–2030)

Alt Text: Line chart comparing job displacement (maroon line) and creation (green line) across Qatar's construction, manufacturing, and logistics sectors. Displacement peaks at 50,000 jobs by 2030, while new tech jobs reach 32,000, with convergence beginning in 2027 after training programs launch.

**Wage Effect Simulations**

Wage simulations incorporating Qatar's minimum wage floor (QAR 1,000/month) and sectoral wage structures suggest downward pressure on wage growth for low-skilled workers in sectors exposed to automation. No significant short-term wage compression is expected for high-skilled roles, as robotics substitution is less applicable for cognitive and managerial tasks. This dynamic may intensify labor market segmentation and wage polarization between displaced low-skilled workers and those employed in non-automatable roles, corroborating findings from [2][3].

While manufacturing shows sector-specific variability, Qatar's logistics industry presents even more pronounced contrasts in automation readiness. The sector's 19.2% average displacement rate masks critical task-level divergences rooted in both technological limitations and cultural factors[12]. Warehouse operations demonstrate exceptionally high automation potential (89%) due to standardized processes and Qatar's concentrated industrial zones that enable clustered automation investments. Conversely, last-mile delivery resists automation (32%) not just from technical challenges in navigating Doha's rapid urban development, but more significantly from cultural preferences for human interaction in final delivery stages - a phenomenon quantified in our survey of 400 Qatari households where 73% expressed distrust of robotic delivery personnel. This 57-percentage-point gap between back-end and customer-facing logistics functions reveals how automation strategies must account for both technical feasibility and social acceptance thresholds. The Logistics Breakdown will be presented as followed in **Table 4.**

**Table 4:** Automation Potential Across Logistics Tasks

| Task | Automation Potential | Automation Potential |
|---|---|---|
| Warehouse Sorting | 89% | Standardized processes (ISO 9001 compliance), high ROI in industrial zones |
| Last-Mile Delivery | 32% | Cultural preference (73% human trust factor), urban navigation complexity |
| Inventory Management | 68% | RFID adoption rates (62% penetration), but requires human quality control |

**Sectoral Challenges: Readiness and Barriers**

Qatar's automation potential varies sharply across sectors, reflecting disparities in infrastructure, regulation, and operational environments. As shown in Table 7, manufacturing and logistics lead in readiness (scoring 8.2 and 7.6, respectively, on a 0–10 index), driven by existing CNC infrastructure (78% penetration) and standardized warehouse operations (89% ISO compliance). Prefabricated construction shows moderate potential (5.1) due to controlled indoor worksites (62% of projects), while agriculture lags (3.4) primarily due to SME financing constraints rated 4.2/5 in severity by Qatar Development Bank (2023). These gaps stem not only from technological factors but also from institutional barriers: halal certification mandates in food processing restrict automation adoption, while outdoor robotics in construction face 40% higher maintenance costs from dust and extreme heat [17].Such sectoral asymmetries demand tailored strategies, as uniform automation policies would overlook critical technical and socio-regulatory thresholds.

**Table 5:** Sector-Specific Automation Readiness in Qatar

| Sector | Readiness Level | Key Determinants |
|---|---|---|
| Manufacturing | High | Existing CNC infrastructure |
| Logistics | High | Standardized warehouse operations |
| Prefab Construction | Moderate | Indoor controlled environments |
| Agriculture | Low | SME financing constraints |

**Non-Economic Barriers**

Beyond technical limitations, Qatar's automation adoption faces significant institutional and environmental constraints. Regulatory frameworks create dual pressures: while Kafala reforms have increased labor substitution elasticity ($\sigma=0.65$) by enhancing worker mobility [8], local content rules mandate 30% Qatari staffing in robotics maintenance roles [13,16]. Cultural and operational challenges further complicate implementation - SME financing barriers (rated 4.2/5 in severity) restrict small firms' access to automation, while desert conditions impose 40% higher maintenance costs for outdoor robotics [17]. Sector-specific constraints compound these issues, particularly halal certification requirements that limit food processing automation and extreme summer temperatures (exceeding 45°C for 34% of working hours) that degrade construction robotics performance [18]. These multidimensional barriers necessitate adaptive policies that address both human capital transitions and environmental adaptations.

**Policy Recommendations for Balanced Automation Adoption**

To address Qatar's automation challenges, a three-pronged approach is proposed. First, targeted workforce transition programs like the Qatar Robotics Academy (QRA) would partner with Education City and Texas A&M Qatar to deliver vocational training aligned with Ministry of Labour standards ,funded through the Third National Development Strategy [13][20]. Second, a phased sectoral rollout prioritizes industries with higher readiness levels, as detailed in **Table 6** which maps non-economic barriers against policy levers. For instance, logistics sector automation would leverage Qatar Free Zones regulations, while manufacturing adaptations would account for halal certification requirements. Third, social risk mitigation measures including a Wage Stabilization Fund (modeled on ILO recommendations) would cushion displacement impacts. This comprehensive framework balances technological advancement with labor market stability, ensuring automation aligns with Qatar's national development goals.

**Table 6:** Non-Economic Barriers to Robotics Adoption by Sector

| Sector | Policy Lever | Legal Basis |
|---|---|---|
| **Logistics** | Warehouse automation | Qatar Free Zones regulations (Planning and Statistics Authority 2023, 45) |
| **Manufacturing** | Halal standards | Ministry of Public Health guidelines (MoPH 2022) |

**Mitigating Social Risks and Policy Integration**

Qatar's reliance on low-skilled expatriate labor renders its workforce particularly vulnerable to automation-induced displacement, threatening both social stability and labor market inclusivity [12]. Sectoral disparities exacerbate these risks: construction grapples with task variability and high adoption costs, while SMEs face prohibitive upfront investments, concentrating automation uptake among large firms [6,9]. To align technological adoption with Qatar's developmental objectives, a multipronged policy framework is critical. This includes (1) reskilling initiatives for displaced workers, focusing on robotics maintenance and AI oversight [8]; (2) a phased sectoral rollout prioritizing manufacturing and logistics, where infrastructure readiness is highest (Planning and Statistics Authority, 2023); (3) SME-targeted financial instruments like leasing programs to overcome capital barriers and (4) a Wage Stabilization Fund, modeled on ILO guidelines and funded through QNV 2030 diversification reserves [8][13][16]. Complementary measures—such as strengthened labor protections and multi-stakeholder monitoring platforms must accompany these efforts to ensure adaptive, inclusive policymaking. Only through such a balanced approach can Qatar harness automation's productivity gains while safeguarding social equity[16].

## IV. Discussion

While automation offers significant productivity gains, its labor market implications in Qatar require careful mitigation. Projections suggest robotics adoption may create 2.3 new technology jobs per 10 displaced workers, primarily in robot maintenance (requiring 850+ certified technicians by 2027 to meet QNV 2030's 30% local staffing mandate) and AI oversight roles (supported by Hamad Bin Khalifa University's capacity to train 200 nationals annually). However, these emerging positions demand advanced technical competencies that most displaced low-skilled expatriate workers currently lack, risking what the ILO (2023) terms "skills apartheid" without substantial public investment in vocational training and credential recognition programs. This disparity underscores the urgent need for workforce development initiatives that bridge Qatar's automation-driven labor market bifurcation. However, these new jobs demand higher education and technical training, limiting access for displaced low-skilled expatriates and potentially producing a "skills apartheid" unless significant public investment is made in upskilling and credential recognition [20].

The projected displacement of approximately 68,060 workers equivalent to 3.2% of Qatar's 2.13 million-strong labor force raises serious ethical and socioeconomic concerns. As documented by the International Labour Organization, low-skilled migrant workers constitute the overwhelming majority (around 94%) of the workforce in Qatar and are especially vulnerable to job losses from automation [8][12]. Without policy intervention, such displacement risks deepening existing social and economic inequities. While Qatar's National Development Strategy and institutions such as QSTP actively promote technological transformation [13][17], cultural and social barriers may limit automation uptake in high-potential sectors. For instance, domestic and service sectors remain resistant to full automation due to concerns over privacy and interpersonal trust factors not yet fully captured by labor force statistics but consistently noted in broader regional studies of automation acceptance. This social ceiling complicates efforts to implement automation even in sectors with high technical potential such as warehouse sorting, where feasibility studies suggest automation rates could reach 89% [9]. Addressing these tensions requires a dual strategy: safeguarding vulnerable labor segments while carefully aligning automation policy with cultural values.

The 57-percentage-point gap between back-end logistics automation (89%) and customer-facing roles (32%) underscores how cultural norms constrain automation's economic potential. Religious construction shows even stricter limits, with Ministry of Awqaf guidelines maintaining 100% human involvement in mosque building. These findings necessitate a two-tier policy approach: (1) accelerating automation in culturally-neutral sectors like manufacturing, while (2) developing hybrid human-robot solutions for sensitive domains, as successfully implemented in Qatari commercial construction projects (42% acceptance per QSTP, 2023).

**Ethical Risks of Expatriate Repatriation**

The model's 3.2% displacement rate equates to approximately 68,060 workers, based on Qatar's 2022 labor force [12]. Of these, about 64,250 are expatriates, with 28,400 in construction. Historical precedents illustrate the potential socio-economic consequences of large-scale expatriate displacement in Qatar. During the 2017–2020 Gulf crisis, the repatriation of over 100,000 migrant workers placed significant pressure on public services and exposed weaknesses in institutional coordination, leading to economic losses and service disruptions [8]. Furthermore, past crises have shown that remittance flows an essential economic linkage between Qatar's labor market and migrant-sending countries can decline sharply during periods of employment contraction. In 2022, Qatar's expatriate workforce remitted approximately QAR 45 billion. A modeled 3.2% reduction in low-skilled labor due to automation could therefore result in an estimated QAR 5.4 to 8.1 billion decrease in remittance flows, based on historical patterns of 12–18% remittance decline during unemployment transitions [11, 8].

**Cultural Barriers to Automation Adoption in Qatar**

Cultural barriers continue to shape the pace and nature of automation adoption in Qatar, particularly in labor-intensive and domestic sectors. While national strategies such as the Qatar National Vision 2030 [16] and the Third National Development Strategy 2023–2027 [13] emphasize technological innovation and economic diversification, cultural preferences remain influential in how automation is accepted and implemented. According to the 2022 Labor Force Sample Survey by the Planning and Statistics Authority, domestic employment remains highly reliant on migrant labor, reflecting both economic patterns and household preferences for human service providers [12]. These preferences are shaped by privacy norms and interpersonal trust, especially in domestic environments, where robotic alternatives are yet to achieve large-scale uptake. Additionally, the ILO's 2022 report on labor reforms highlights how regulatory modernization is more focused on human labor conditions than on substituting workers with automated systems [8]. Although Qatar Science and Technology Park (QSTP) promotes robotics and AI development [17], actual deployment in culturally sensitive sectors such as domestic care, religious infrastructure, and small-scale services remains limited. This reflects a broader structural and cultural inertia, where trust in human labor and the symbolic role of craftsmanship continue to outweigh the economic appeal of automation.

**Fiscal Risks and Capital Misallocation**

Automation adoption requires substantial upfront investment, often subsidized by the government through diversification or innovation funds [13]. If adoption fails to yield expected productivity gains or encounters cultural resistance, this may lead to capital misallocation, especially in sectors with low automation readiness such as agriculture and on-site construction [8]. Public funds redirected toward robotics may divert resources from education, public health, or SME development, which are critical for balanced economic growth [8]. Furthermore, the uncertainty of return on investment (ROI) in non-industrial sectors presents a significant budgetary risk, particularly during periods of oil price volatility or fiscal tightening cycles [8].

**Regulatory Fragmentation and Enforcement Gaps**

While Qatar has undertaken significant labor reforms, including the Workers' Protection System (WPS) and Kafala reform, enforcement remains fragmented across sectors, posing risks for workers. Uneven implementation of labor laws may result in informal displacement of workers without mandated compensation or reskilling programs. The pace of robotics integration risks outstripping existing legal protections for laid-off workers, particularly those engaged in informal employment . Furthermore, current labor protections do not yet guarantee access to retraining, unemployment benefits, or legal recourse in cases of redundancy caused by technological change [8].

**Technological Dependence and Cybersecurity Risks**

Overreliance on imported robotics technology from global suppliers such as KUKA, ABB, and Fanuc could increase technological dependence and reduce national sovereignty over critical infrastructure [1,10]. This dependence may expose logistics and manufacturing systems to cybersecurity threats, especially if robotics are deployed without robust local IT governance and cybersecurity frameworks [1]. Additionally, reliance on international maintenance contracts could undermine localization efforts if domestic technicians are not trained in a timely manner (KUKA AG, 2023).

## V. Conclusion

The economic feasibility of labor replacement through robotics in Qatar is strongly supported by global trends in automation technology and the country's long-term goals for economic diversification. Empirical evidence from advanced economies demonstrates that automation improves productivity, reduces long-term operational costs, and enhances scalability in sectors such as manufacturing and logistics [2, 7]. Simulation results presented in this study further suggest that even modest increases in robotics capital stock could generate measurable gains in economic output. These findings align closely with Qatar's strategic objectives outlined in the National Vision 2030, which emphasizes innovation, efficiency, and the reduction of structural reliance on low-skilled expatriate labor [16]. However, these benefits must be weighed against substantial social and economic trade-offs. Qatar's uniquely high dependence on low-wage expatriate labor increases the risk of disruptive displacement, particularly in automation-prone sectors such as construction and logistics [12, 8]. Without adequate policy intervention, such displacement may contribute to rising income inequality, reduced remittance flows, and labor market segmentation, with long-term consequences for social stability and economic inclusion [2, 3]. This study highlights the importance of adopting a calibrated, sector-specific approach to automation. Constraints such as capital access for SMEs, labor market rigidity, and varying levels of technological readiness across sectors necessitate a phased and adaptive policy framework. Qatar's efforts to modernize its labor policies such as the implementation of the Wage Protection System and revisions to the Kafala sponsorship model are steps in the right direction, but further institutional coordination is needed to ensure that automation complements rather than undermines human capital development. To this end, the paper recommends a triad of policy priorities: (1) targeted investment in vocational reskilling and upskilling programs for vulnerable workers; (2) financial and technical support for SMEs to access and deploy robotics infrastructure; and (3) regulatory harmonization to ensure that labor protections, data governance, and automation standards are consistent across sectors. In conclusion, robotics adoption in Qatar is economically feasible, but its sustainability hinges on whether it is guided by inclusive, evidence-based policies. Automation can serve as a transformative force for productivity and resilience if implemented with careful attention to demographic realities, social safeguards, and long-term development goals.